\documentclass[lettersize,journal]{IEEEtran}
\usepackage{amsmath,amsfonts}
\usepackage{algorithmic}
\usepackage{algorithm}
\usepackage{array}
\usepackage[caption=false,font=normalsize,labelfont=sf,textfont=sf]{subfig}
\usepackage{textcomp}
\usepackage{stfloats}
\usepackage{url}
\usepackage{verbatim}
\usepackage{makecell}
\usepackage{array}
\usepackage{hyperref}
\usepackage{graphicx}
\usepackage{cite}
\usepackage{multirow}
\usepackage{subfloat}
\usepackage{pifont}
\usepackage{xcolor}
\begin{document}

\title{Learning-Based Hybrid Neural Receiver for 6G-V2X Communications}
\author{Osama Saleem, Mohammed Alfaqawi, Pierre Merdrignac, Abdelaziz Bensrhair, Soheyb Ribouh
\thanks{Osama Saleem, Abdelaziz Bensrhair and Soheyb Ribouh are with INSA Rouen Normandie, Univ Rouen Normandie, Université Le Havre Normandie, Normandie Univ, LITIS UR 4108, Rouen, France. Mohammed Alfaqawi and Pierre Merdrignac are with Institut VEDECOM, Versailles, France. (emails: \{osama.saleem, abdelaziz.bensrhair\}@insa-rouen.fr; \{mohammed.alfaqawi, pierre.merdrignac\}@vedecom.fr;  soheyb.ribouh@univ-rouen.fr)}
\thanks{The corresponding author is Osama Saleem}}


\maketitle

\begin{abstract}
Neural receiver models are proposed to jointly optimize multiple functionalities of wireless receivers; however, a comprehensive receiver model that replaces the entire physical layer blocks has not yet been presented in the literature. In this work, we introduce a novel hybrid neural receiver (H-NR) built on Transformer encoder blocks and Graph Neural Network (GNN), as part of an end-to-end wireless communication framework. In our communication framework, we assume vehicle to network (V2N) uplink scenario where information is transmitted by vehicle and received at the base station (BS). Our proposed H-NR model replace OFDM resource grid demapping, channel estimation, signal equalization, demodulation, and channel decoding. To test the adaptability of our proposed model on unseen conditions, we evaluate its performance for various scenarios, including a vehicle speed of range [$0-60$] km/h, a carrier frequency of $5.9$GHz, and a cluster delay line (CDL) channel model. Furthermore, we assess the performance of our proposed H-NR on multimodal data, such as images, audio, GPS, radar, and LiDAR, to examine its adaptability in real-world use cases. 
The simulation results clearly demonstrate that our proposed model outperforms the state-of-the-art neural receiver by approximately $0.5$ dB in terms of reconstruction and error correction.
\end{abstract}


\section{Introduction} \label{s1}
\IEEEPARstart{T}{he} emergence of sixth-generation (6G) wireless networks is expected to revolutionize global communication infrastructures through unprecedented advancements in system design. Emerging technologies such as massive multiple input multiple output (MIMO), millimeter wave (mmWave) spectrum utilization, and reconfigurable intelligent surfaces (RIS) are expected to substantially improve network capabilities \cite{b3}. They enable ultra-high data rates, near-real-time latency, and enhanced resilience against evolving cyber threats. The primary importance of these advancements can be attributed to the emergence of Artificial Intelligence (AI)-driven solutions. These solutions play a crucial role in optimizing physical-layer protocols and enabling intelligent and adaptive resource management. This transformative approach is in line with the strategic vision of the 3rd Generation Partnership Project (3GPP), which has officially recognized AI-driven techniques as essential components in the evolution of future wireless standards \cite{b1}.

\begin{figure}[t]
  \centering
  \includegraphics[width=0.45\textwidth,height=0.25\textheight]{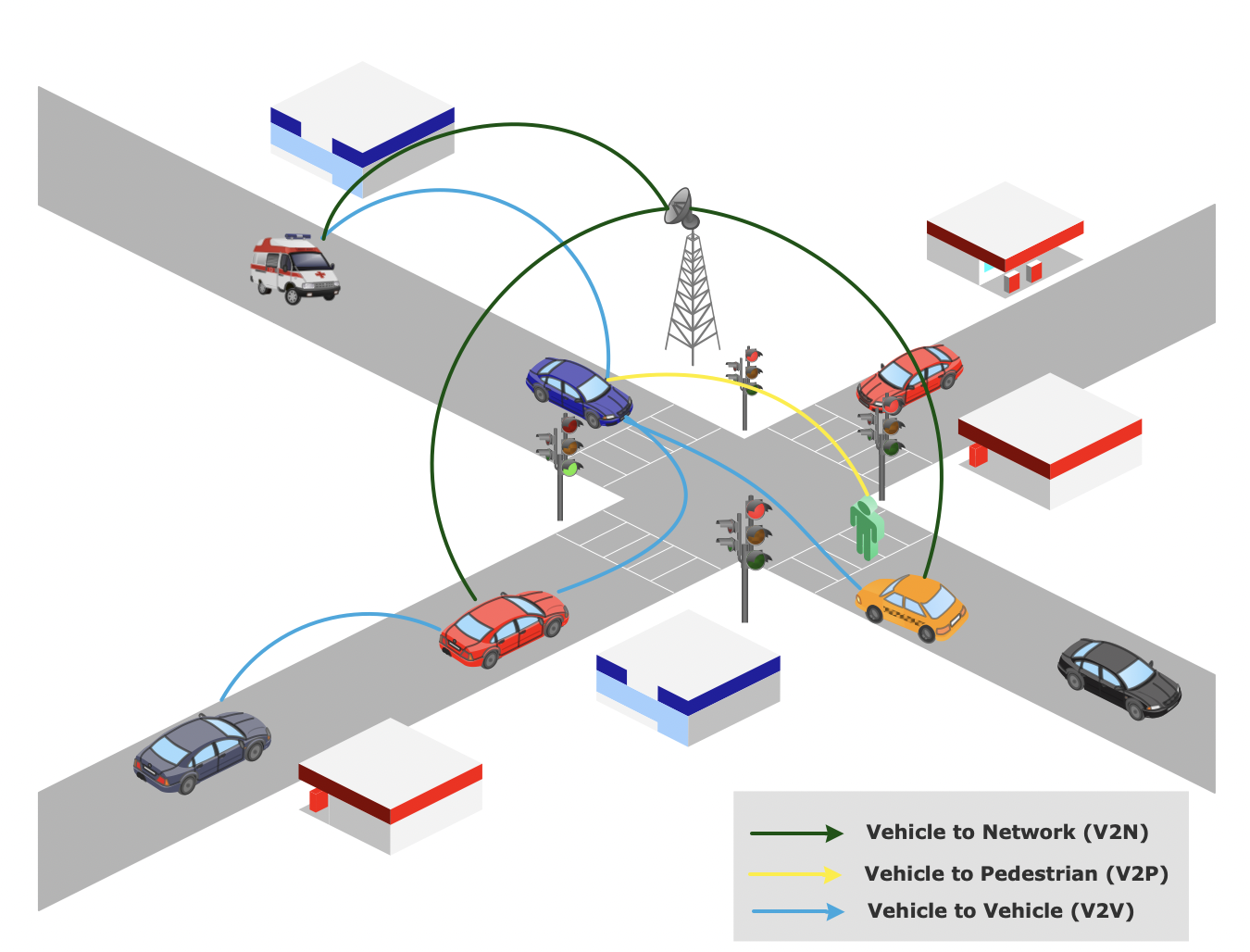}
  \caption{Vehicle-to-Everything (V2X) Communication in Urban Environment}
  \label{fig1}
\end{figure}

A critical application domain for 6G lies in cooperative connected and autonomous mobility (CCAM) systems \cite{b2}, which form the backbone of intelligent transportation networks that rely on autonomous vehicles (AVs). Current AVs integrate multimodal sensor data-including vision-based cameras, radar systems, and light detection and ranging (LiDAR) units to perpetually acquire high-resolution environmental data \cite{b4}. This data is shared across vehicular ad-hoc networks via vehicle-to-everything (V2X) communication as shown in Fig.~\ref{fig1}. It helps to facilite the AV for collaborative perception and real-time trajectory planning. These interconnected environments require a need for robust and reliable information reconstruction mechanisms to ensure operational safety and decision-making accuracy in dynamic environments.

Recent trends emphasize the importance of developing AI-based intelligent and adaptive reconstruction techniques to recover multimodal sensor data that is degraded by the wireless channel impairments. In such contexts, traditional Physical Layer (PHY) methods may fall short, especially when dealing with sparse and noisy inputs. As a result, Large AI model-based frameworks are increasingly being explored for their potential to learn robust mappings from transmitted signals and accurately reconstruct information bits in future wireless communication systems. By integrating near-real channel characteristics such as path loss and shadow fading into the learning process, such models are capable of dynamically adjusting their reconstruction strategy in response to the communication environment. This context sensitivity enables them to maintain high reconstruction accuracy under varying vehicular mobility and network conditions, which are inherent in CCAM scenarios.

\subsection{Contributions}
Our main contributions include the following:

\begin{enumerate}
    \item We propose a novel hybrid neural receiver model, called H-NR. Our proposed model is built on Transformer encoder blocks and a Graph Neural Network (GNN). It replaces traditional receiver functions including resource grid demapping, channel estimation, equalization, demodulation, and channel decoding at the receiver. The transformer blocks handle the signal detection and Log Likelihood Ratio (LLR) computation, while a bipartite GNN decodes the signal using a message passing method between the nodes. The model takes a received resource grid as input and outputs reconstructed information bits.

    \item We evaluate the adaptation capabilities of our proposed H-NR model-based end to end communication framework by evaluating it across diverse scenarios other than the one used during training. It includes testing it on low frequency of $5.9$GHz, AV speed range of $0–60$km/h, and Cluster Delay Line (CDL) channel models. Result from simulation shows that our proposed H-NR model maintains consistent performance across all conditions without requiring re-training.

    \item We also evaluate the multimodal processing capabilities of our proposed H-NR model using various AV sensor data including image, audio, GPS, radar, and LiDAR. These data streams are transmitted over a realistic simulated wireless channel, and reconstructed at the base station (BS) by our proposed H-NR.

\end{enumerate}

\subsection{Organization}
The rest of this paper is organized as follows: Section~\ref{s2} presents a systematic review of the literature on AI-based receiver architectures. Section~\ref{s3} introduces the baseline receiver architecture formulated using stochastic equations. Section~\ref{s4} details our proposed Hybrid Neural Receiver (H-NR) architecture and its training methodology. Section~\ref{s5} specifies the experimental setup and implementation framework for testing the H-NR. Section~\ref{s6} presents the empirical evaluation results in various V2X use cases and application scenarios. Finally, Section~\ref{s7} concludes with key findings and outlines future research directions in the domain.

\section{Related Works} \label{s2}
The concept of an autoencoder-based end-to-end (E2E) wireless communication system was first introduced in \cite{c1}. The primary goal was to leverage deep learning techniques to model and detect the nonlinear imperfections present in real-world wireless channels. Although their initial work focused on a simplified communication scenario, it marked a significant advancement for physical layer design in next-generation wireless systems. Since then, numerous algorithms have been developed to optimize transmitter and receiver operations using deep learning.

\begin{table*}[t]
\caption{Comparison of proposed Full Receiver Substitution (\textbf{FRS}) Method, H-NR, with state of the art Single Module Substitution (\textbf{SMS}) and Multiple Module Substitution (\textbf{MMS}) approaches.}
\centering
\resizebox{\textwidth}{!}{%
\begin{tabular}{|>{\centering\arraybackslash}m{3.2cm}||>{\centering\arraybackslash}m{4.5cm}||c||c||c||c||c|}
\hline
\textbf{Reference} & \textbf{Method} & \textbf{Resource Grid Demapper} & \textbf{Estimation} & \textbf{Equalization} & \textbf{Demodulation}  & \textbf{Channel Decoding} \\
\hline
\cite{c2}, \cite{c3} & SMS: Estimation & \ding{55} & \ding{51} & \ding{55} & \ding{55} & \ding{55}\\
\hline
\cite{c4} & SMS: Equalization & \ding{55} & \ding{55} & \ding{51} & \ding{55} & \ding{55} \\
\hline
\cite{c5}, \cite{c81} & SMS: Demodulation & \ding{55} & \ding{55} & \ding{55} & \ding{51} & \ding{55} \\
\hline
\cite{c82} & SMS: Decoder & \ding{55} & \ding{55} & \ding{55} & \ding{55} & \ding{51} \\
\hline
\makecell[c]{\cite{c9}, \cite{c10}, \cite{c11}, \cite{c111},\\ \cite{c112}, \cite{c113}, \cite{c114}} & MMS: Neural Receiver & \ding{55} & \ding{51} & \ding{51} & \ding{51} & \ding{55} \\
\hline
Proposed & FRS: H-NR & \ding{51} & \ding{51} & \ding{51} & \ding{51} & \ding{51} \\
\hline
\end{tabular}%
}
\label{table1}
\end{table*}

In \cite{c2}, the authors utilize a machine learning model for channel estimation. The channel vector in conventional communication is stochastic and depends on random parameters. If the covariance matrices exhibit shift invariance structures, the computational complexity of the Minimum Mean Squared Error (MMSE) channel estimator can be reduced to $O(M \log M)$, where $M$ denotes the channel dimension. However, in the absence of shift invariance, the complexity becomes significantly higher. This work proposes a Neural Network (NN)-based MMSE estimator for the unstructured model, achieving complexity comparable to that of the covariance matrix with shift invariance. However, in millimeter wave (mmWave) multiple-input multiple-output (MIMO) communication, when the receiver is equipped with a limited number of radio-frequency (RF) chains, channel estimation becomes even more challenging. To address this issue, He \emph{et al.} \cite{c3} propose a learned denoising-based approximate message passing (LDAMP) network. This approach outperforms stochastic compressed sensing techniques. To solve the problem of signal equalization, Chang \emph{et al.} \cite{c4} propose a complex-valued convolutional neural network (CNN) to recover the transmitted signal from the received channel data. To evaluate the generalization capability of the proposed algorithm, they generate a noisy dataset containing known modulation signals, taking multipath fading into consideration. Their results demonstrate that the proposed method outperforms stochastic equalization techniques such as the multi-modulus algorithm, least mean squares method, and recursive least squares method. In the field of demapping, Shental \emph{et al.} \cite{c5} introduce a deep neural network framework aimed at efficiently calculating bit log-likelihood ratios (LLRs) for equalized symbols. This deep learning-based demapper achieves performance comparable to the optimal log maximum a-posteriori algorithm, while significantly reducing computational complexity. Additionally, to address the complexity associated with machine learning (ML)-based solutions for channel estimation and equalization, Michon \emph{et al.} \cite{c81} proposed a convolutional self-attention-based MIMO demapper for multi-user systems. Their work employs a simplified channel estimator, providing an imperfect signal as input to the proposed demapper, which still outperforms the stochastic baseline. Furthermore, Cammerer \emph{et al.} \cite{c82} propose a fully differentiable, message-passing-based Graph Neural Network (GNN) for channel decoding. Their approach considers received LLRs as variable nodes and parity bits as check nodes, with connections established between them through a parity-check matrix (PCM). Their work shows that the GNN-based decoder outperforms the Bose–Chaudhuri–Hocquenghem (BCH) decoder and achieves performance comparable to that of the low density parity check (LDPC) decoder. Moreover, several methods investigate how deep learning (DL) can be incorporated into the standard receiver processing \cite{c6,c7,c8}. These DL based improved systems perform better than conventional stochastic receiver when properly trained. However, there optimization is limited to a single receiver functionality only.

To optimize multiple receiver functionalities, Ye \emph{et al.} \cite{c9} propose a deep learning-based method that jointly optimizes channel estimation and signal detection. Their approach utilizes a fully connected neural network that analyzes both pilot and data signals, which demonstrates significant performance improvements compared to traditional MMSE-based stochastic receivers, especially in scenarios with limited channel estimation pilots or in the absence of a cyclic prefix. In contrast, Zhao \emph{et al.} \cite{c10} employ CNN to develop a receiver that directly estimates bits from the received time-domain signal. This approach demonstrates enhanced performance, particularly in low to medium signal-to-noise ratio (SNR) scenarios. While the CNN-based receiver maintains an advantage over linear least squares-based receivers even at higher SNRs, its accuracy falls short of that achieved by MMSE-based receivers with perfect channel knowledge. Honkala \emph{et al.} \cite{c11} propose a fully CNN-based deep neural receiver architecture designed to process 5th Generation (5G) signals. This approach achieves superior performance over traditional stochastic algorithms by leveraging data and pilot symbols for precise channel estimation and generating soft bits compatible with 5G channel coding. Furthermore, Pihlajasalo \emph{et al.} \cite{c111} propose a hybrid variant of DeepRx \cite{c11} called HybridDeepRx, which utilizes Orthogonal Frequency Division Multiplexing (OFDM) signals as input. They introduce a dual-domain CNN architecture that processes the signals in both time and frequency domains. This approach enables robust demodulation and decoding, even under conditions of high error vector magnitude (EVM) resulting from power amplifier (PA) saturation non-linearity. In order to further improve signal detection for Additive White Gaussian Noise (AWGN) channel, Çevik \emph{et al.} \cite{c114} proposed the Dual Attention Transformer (DAT) and the Residual Dual Non-Local Attention Network (RDNLA), which combine residual learning and a dual self-attention mechanism. These architectures are intended to reduce interference and maximize feature representation in complex communication environments.


Although the research works presented in \cite{c9,c10,c11,c111} propose joint optimization of the neural receiver, their solution lacked domain generalization (DG) \cite{c12}, which concerns the adaptability of a model when exposed to out-of-distribution (OOD) data i.e., different from the training data. This problem is addressed by \cite{c13,c112,c113} which train the model on urban channel. Cammerer \emph{et al.} \cite{c13} propose a Neural Receiver model for a 5G NR multi-user MIMO system. Their system utilizes a CNN-based architecture to compute soft bits and a GNN-based architecture to manage the multi-user scenario. Their model adapts to varying numbers of subcarriers and supports multiple MIMO layers without requiring retraining. Furthermore, Saleem \emph{et al.} \cite{c112} proposes a transformer-based architecture named TransRx, designed for 6G V2X communication. Their system takes the received resource grid as input and outputs optimal LLRs. Their proposed model integrates and replaces the channel estimation, signal equalization, and demodulation functions of the receiver architecture, demonstrating superior end-to-end performance compared to CNN-based receivers. Furthermore, in \cite{c113}, the authors use an attention mechanism and investigate the impact of varying training parameters to optimize the performance in term of BER. The model's multimodal capabilities are validated across diverse V2X data, showing improved performance compared to state-of-the-art receivers.  While previous work has focused on neural receiver models that outperform conventional stochastic receiver architectures, their optimization is limited to multiple modules substitution (MMS) as shown in Table~\ref{table1}. However, in this work, we propose a full receiver substitution (FRS) architecture, called H-NR, that replaces all the functionalities of the stochastic receiver, including resource grid demapping, channel estimator, signal equalization, signal demodulation, and channel decoder. The input to our proposed H-NR model is the received resource grid and the output is reconstructed information bits. To the best of our knowledge, this is the first work that involves the proposition of a full receiver substitution.

\section{System Model} \label{s3}
\begin{figure*}[htb!]
  \centering
  \includegraphics[width=0.85\textwidth,height=0.55\textheight]{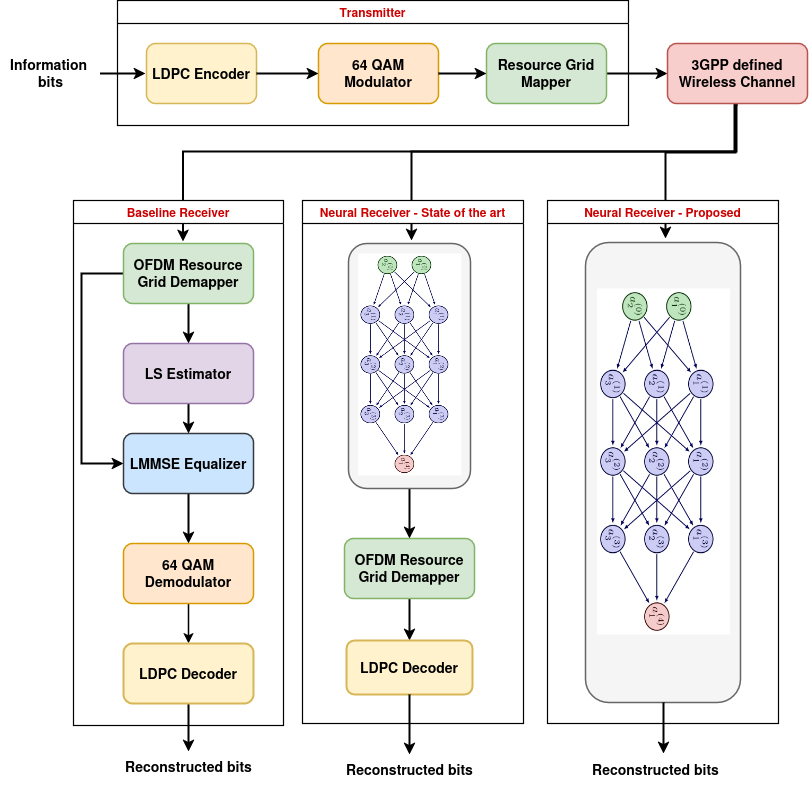}
  \caption{PHY of Proposed H-NR model and state of the art methods based end to end wireless communication framework.}
  \label{fig3}
\end{figure*}

To model our system, we assume a transmitter that processes the information bits through various stages, i.e., channel coding, modulation, OFDM symbol mapping, and transmission over the wireless channel.\\
\textbf{The channel coding} process transforms the information bit sequence into a codewords using:
\begin{equation}
\mathbf{c} = \mathcal{C}(\mathbf{b}, G)    
\end{equation}
Where:
\begin{itemize}
    \item $\mathbf{b} = [b_1, b_2, \dots, b_K] \in \{0, 1\}^K$ is the information bit sequence of length $K$,
    \item $\mathcal{C}(\mathbf{b}, G)$ represents the channel coding function (e.g., LDPC),
    \item $G$ is the generator matrix,
    \item $\mathbf{c} = [c_1, c_2, \dots, c_N] \in \{0, 1\}^N$ is the coded bit sequence of length $N$, where $N > K$.
\end{itemize}
These encoded bit sequence is then \textbf{mapped} onto higher order modulation symbols as shown in Fig.~\ref{fig3}. The modulation function is described as:
\begin{equation}
\mathbf{x} = \mathcal{M}(\mathbf{c})    
\end{equation}
Where:
\begin{itemize}
    \item $\mathcal{M}(\mathbf{c})$ represents the modulation function (e.g., QPSK, 16-QAM, 32-QAM, 64-QAM),
    \item $\mathbf{x} = [x_1, x_2, \dots, x_M] \in \mathbb{C}^M$ is the modulated symbol sequence,
    \item $M = \frac{N}{\log_2(Q)}$ is the length of the modulated symbol sequence,
    \item $Q$ is the modulation order, where $Q = 4$ for QPSK, $Q = 16$ for 16-QAM and $Q = 32$ for 32-QAM etc.
\end{itemize}
The modulated symbols are then mapped onto an \textbf{OFDM resource grid}, with the Inverse Fast Fourier Transform (IFFT) applied to convert them into time-domain symbols.
\begin{equation}
\mathbf{X}[k, l] = \mathcal{F}(\mathbf{x}[k, l])    
\end{equation}
Where:
\begin{itemize}
    \item $\mathbf{X}[k, l] \in \mathbb{C}$ represents the modulated symbols mapped to subcarrier $k$ and OFDM symbol $l$,
    \item $\mathcal{F}(\mathbf{x}[k, l])$ is the IFFT operation on the modulated data,
    \item $k$ is the subcarrier index and $l$ is the OFDM symbol index.
\end{itemize}
\textbf{The channel model} used for the vehicular scenario is based on the UMa model defined by 3GPP \cite{c115}. \\

\textbf{The received signal} for subcarrier $k$ and OFDM symbol $l$ can be expressed as:
\begin{equation}
\mathbf{Y}[k,l] = \mathbf{H}[k,l] \mathbf{X}[k,l] + \mathbf{N}[k,l]
\end{equation}
Where:
\begin{itemize}
    \item $\mathbf{Y}[k,l] \in \mathbb{C}^{N_r \times 1}$ is the received signal vector,
    \item $\mathbf{H}[k,l] \in \mathbb{C}^{N_r \times N_t}$ is the frequency-domain channel matrix,
    \item $\mathbf{X}[k,l] \in \mathbb{C}^{N_t \times 1}$ is the transmitted symbol vector,
    \item $\mathbf{N}[k,l] \sim \mathcal{CN}(0, \sigma_n^2 \mathbf{I})$ is the noise vector.
\end{itemize}
Receiver first \textbf{estimates the channel} using:
\begin{equation}
\hat{\mathbf{H}}[k,l] = \mathbf{Y}[k,l] \mathbf{P}^H[k,l] (\mathbf{P}[k,l] \mathbf{P}^H[k,l])^{-1}
\end{equation}
Where:
\begin{itemize}
    \item $\mathbf{P}[k,l]$ is the pilot symbols matrix,
    \item $^H$ denotes the Hermitian (conjugate transpose) operation,
    \item $\tilde{\mathbf{H}}[k,l]$ is the estimated channel vector.
\end{itemize}
In the next step, \textbf{equalization} is performed on the received signal to mitigate multipath fading and inter-symbol interference using:
\begin{equation}
\hat{\mathbf{X}}[k,l] = \left( \hat{\mathbf{H}}^H[k,l] \hat{\mathbf{H}}[k,l] + \sigma_{\text{eff}}^2[k,l] \mathbf{I} \right)^{-1} \hat{\mathbf{H}}^H[k,l] \mathbf{Y}[k,l]
\end{equation}
Where:
\begin{itemize}
    \item $\sigma_\text{eff}^2$ is the effective noise,
    \item $\mathbf{I}$ is the identity matrix.
\end{itemize}
For M-QAM modulation (M = $4$, $16$, $32$ etc) and bit $b_i$, the received signal is \textbf{demodulated} to compute LLR as:
\begin{equation}
LLR(b_i) = \log \left( \frac{P(b_i = 1 | \mathbf{\hat{X}}, \mathbf{H})}{P(b_i = 0 | \mathbf{\hat{X}}, \mathbf{H})} \right)
\end{equation}
Where:
\begin{itemize}
    \item $P(b_i = 1 | \mathbf{\hat{X}}, \mathbf{H})$ is the probability that the received bit is $1$,
    \item $P(b_i = 0 | \mathbf{\hat{X}}, \mathbf{H})$ is the probability that the received bit is $0$
\end{itemize}
At the last step, \textbf{decoding} is performed using belief propagation (BP) with $10$ message passing iterations to reconstruct original message using:
\begin{equation}
\mathbf{\hat{b}} = \arg \min_{\mathbf{b}} \sum_{i=1}^{N_b} \left( b_i \cdot LLR(b_i) \right)    
\end{equation}
Where:
\begin{itemize}
    \item $\mathbf{\hat{b}}$ is the reconstructed bit,
    \item $N_b$ is the total number of bits transmitted.
\end{itemize}

\section{Proposed H-NR Framework} \label{s4}
\begin{figure}[b]
  \centering
  \includegraphics[width=0.47\textwidth,height=0.13\textheight]{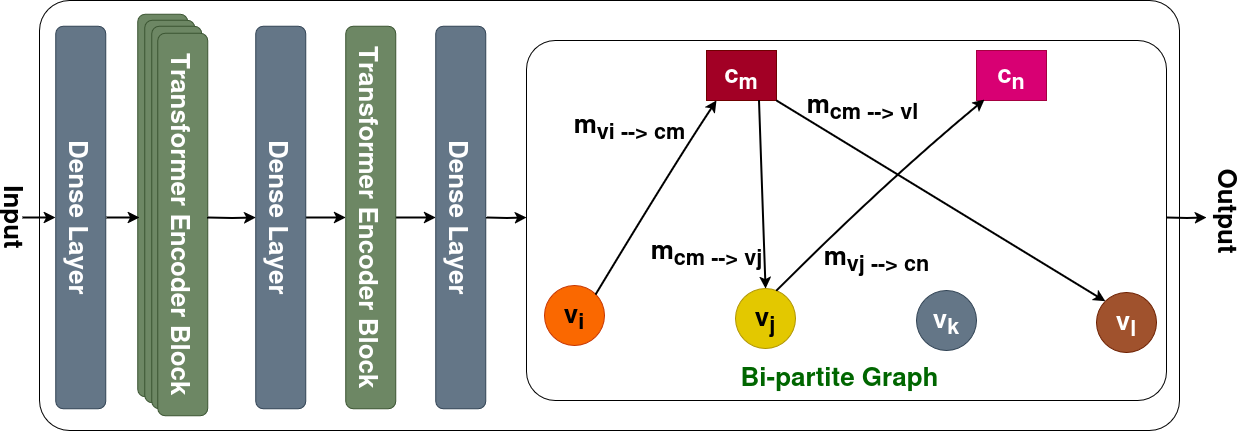}
  \caption{H-NR Architecture}
  \label{fig4}
\end{figure}

We propose a novel hybrid neural receiver model, called H-NR, that is designed to replace and enhance the core functionalities of a baseline PHY of a receiver using deep learning techniques. To train our proposed H-NR model, we take the AV speed range of [$60-120$]km/h with signal transmitted over UMa channel and carrier frequency of $28$GHz. At the BS, the received signal is processed by our proposed H-NR model that replaces OFDM resource grid demapping, channel estimation, signal equalization, higher-order demodulation, and channel decoding. It is trained to accurately reconstruct the transmitted data from the received signal. Our proposed H-NR model takes noisy OFDM signal along with the power of noise introduced in the channel as input and generates the reconstructed bits at the output. The architecture begins with an input dense layer that processes the incoming data and maps them into an appropriate feature space. This is followed by several transformer encoder blocks, which constitute the core computational engine of our proposed model. In total, our architecture has $5$ transformer encoder blocks as mentioned in Table.~\ref{table3}. The architecture of transformer encoder block is same as presented in \cite{c112}. Each Transformer block is built around the multi-head self-attention mechanism, which enables the model to capture long-range dependencies and complex signal patterns within the input data. The attention mechanism is mathematically defined as:

\begin{equation}
\text{Attention}(Q, K, V) = \text{softmax}\left( \frac{QK^\top}{\sqrt{v_k}} \right)V
\end{equation}

where:
\begin{itemize}
    \item $Q$, $K$, and $V$ represent the query, key, and value matrices derived from the input,
    \item $v_k$ is the dimension of the key vectors used to normalize the dot product to ensure stable gradients.
\end{itemize}
In the multi-head attention approach, multiple such attention operations are conducted in parallel, each with different learned projections. These parallel outputs are then concatenated and passed through a final dense layer to generate a unified representation. This allows the model to simultaneously analyze the input data from multiple perspectives to capture diverse signal characteristics at various spatial and temporal scales. Once the data passes through all the transformer encoder layers, it is processed by a final dense layer to yield the predicted LLRs. We trained the transformer encoder blocks and dense layers using the LDPC decoder for error correction on $12$ Million data samples by employing the AdamW optimizer as mentioned in Table. \ref{table3}.

\begin{table}[b]
    \centering
    \caption{H-NR Parameters}
    \begin{tabular}{|c|c|} \hline
        \textbf{Parameter} & \textbf{Value} \\ \hline
        No. of Transformer Blocks & 5 \\ \hline
        No. of Attention Heads & 4 \\ \hline
        Feed Forward Network Dimension & 128 \\ \hline
        Embedding Dimension & 128 \\ \hline
        Transformer Training Data & 12 M \\ \hline
        Transformer Training Learning Rate & $1e^{-4}$ \\ \hline
        Transformer Training Optimizer & AdamW \\ \hline
        GNN Embedding Dimensions & 16 \\ \hline
        GNN Message Dimensions & 16 \\ \hline
        GNN Hidden Units & 48 \\ \hline
        MLP Layers at CN & 3 \\ \hline
        MP Iterations between VN and CN & 12 \\ \hline
        GNN Training Learning Rate & [$5e^{-4}$, $1e^{-4}$, $1e^{-5}$]  \\ \hline
        GNN Training data & [$4$M, $40$M, $40$M] \\ \hline
        GNN and Finetune Optimizer & Adam \\ \hline
        Finetune Learning Rate & $1e^{-5}$ \\ \hline
        Finetune data & 1 M \\ \hline
        Activation Function & Relu \\ \hline
        
    \end{tabular}  
    \label{table3}
\end{table} 

In the next step, we freeze the Transformer block and dense layers and replace the LDPC decoder by a GNN-based channel decoder. We construct a bipartite graph structure using the parity check matrix ($\mathbf{H}$). The received LLRs serve as variable nodes ($v_n$), while the parity bits act as check nodes ($c_n$). Connections (edges) are established between $v_n$ and $c_n$ using $\mathbf{H}$, as illustrated in Fig.~\ref{fig4}. We train a GNN based channel decoder using multiple learning rates of [$5e^{-4}$, $1e^{-4}$, $1e^{-5}$] for [$4$M, $40$M, $40$M] data respectively using Adam as optimizer as mentioned in Table.~\ref{table3}. For each set of codewords during training, we initialize the value for $c_n$ as $'0'$ which is updated after each iteration. We implement the decoding operation over bipartite factor graphs by adopting the Message Passing Neural Network (MPNN) framework \cite{MPNN}. We define two categories of trainable operations in our proposed model, \textbf{(i)} \textit{Edge-level message computation functions}, and \textbf{(ii)} \textit{Node-level embedding update functions}. 

The message from a variable node $v_k$ to a check node $c_m$ is computed as:
\begin{equation}
\textit{m}_{v_k \rightarrow c_m} = \phi_{\text{edge}}^v\left([z_{v_k} \,\|\, z_{c_m} \,\|\, \gamma_{v_k \rightarrow c_m}], \Theta_{v \rightarrow c} \right)
\end{equation}

where:
\begin{itemize}
    \item $z_{v_k}, z_{c_m}$ are the embeddings of nodes $v_k$ and $c_m$,
    \item $\gamma_{v_k \rightarrow c_m}$ is the edge-specific feature vector,
    \item $\Theta_{v \rightarrow c}$ denotes shared trainable weights,
    \item $\phi_{\text{edge}}^v$ is the neural function computing messages from $v_k$ to $c_m$,
    \item $\|$ indicates vector concatenation.
\end{itemize}

The reverse message from check node $c_m$ to variable node $v_k$ is given as:

\begin{equation}
\textit{m}_{c_m \rightarrow v_k} = \phi_{\text{edge}}^c\left([z_{c_m} \,\|\, z_{v_k} \,\|\, \gamma_{c_m \rightarrow v_k}], \Theta_{c \rightarrow v} \right)
\end{equation}

where $\phi_{\text{edge}}^c$ and $\Theta_{c \rightarrow v}$ are the analogous function and parameter set for this direction.

Node embeddings are updated using aggregated incoming messages. For a variable node $v_k$, the update rule is:

\begin{equation}
z'_{v_k} = \phi_{\text{node}}^v\left([z_{v_k} \,\|\, \bigoplus\limits_{c_m \in \mathcal{N}(v_k)} \textit{m}_{c_m \rightarrow v_k} \,\|\, \delta_{v_k}], \Theta_v \right)
\end{equation}

For a check node $c_m$, the update is:

\begin{equation}
z'_{c_m} = \phi_{\text{node}}^c\left([z_{c_m} \,\|\, \bigoplus\limits_{v_k \in \mathcal{N}(c_m)} \textit{m}_{v_k \rightarrow c_m} \,\|\, \delta_{c_m}], \Theta_c \right)
\end{equation}

where:
\begin{itemize}
    \item $\delta_{v_k}, \delta_{c_m}$ are node-specific feature vectors,
    \item $\Theta_v, \Theta_c$ are shared parameters,
    \item $\bigoplus$ is an aggregation operator (sum),
    \item $\mathcal{N}(v_k), \mathcal{N}(c_m)$ denote the neighborhoods of the nodes.
\end{itemize}

Finally, we fine-tune our proposed H-NR neural receiver model, including transformer encoder blocks, dense layers, and a bi-partite graph using learning rate of $1e^{-5}$ for 1 Million data and Adam optimizer. The joint optimization for this step can be written as:

\begin{equation}
\hat{\boldsymbol{\theta}} = \arg \min_{\boldsymbol{\theta}} \left\{ \frac{1}{N_b} \sum_{i=1}^{N_b} \mathcal{L}\left( b_i, \hat{b}_i(\boldsymbol{\theta}) \right) \right\}
\end{equation}

Where:
\begin{itemize}
    \item \( \hat{b}_i(\boldsymbol{\theta}) \) is the soft output (probability) predicted by the neural network for bit \( b_i \),
    \item \( \mathcal{L}\) is the Binary Cross Entropy (BCE) loss function.
\end{itemize}

The gradient of BER with respect to the receiver parameters $\mathbf{\theta}$ is given by:

\begin{equation}
\nabla_{\mathbf{\theta}} \text{BER} = \frac{\partial}{\partial \mathbf{\theta}} \left( \frac{1}{N_b} \sum_{i=1}^{N_b} P(\hat{b}_i \neq b_i; \mathbf{\theta}) \right)
\end{equation}

\section{Experimental Setup and Implementation} \label{s5}
We design a 6G simulated environment using Sionna \cite{c14}. We model an urban environment where information is shared between the AV and the BS in a V2N manner. For the testing of our proposed H-NR model, we consider V2N uplink communication, where the information bits are transmitted by the AV and received at the BS. 
To evaluate the performance of our proposed H-NR model, we conduct experiments in two stages: first, by varying the PHY parameters that impact the channel behavior, and second, by evaluating it on the system level using data involving all the V2X use cases. The following is a detailed explanation of the experimental setup for both analysis. \\
\textbf{Channel Level Setup:} We assume a wireless communication system that utilizes LDPC channel encoding. The encoded bit streams then undergo higher order $64$-QAM modulation to generate complex-valued baseband symbols. Each of these symbols carries $6$ information bits $(2^6 = 64)$. These symbols are mapped onto an OFDM resource grid that contains 129 subcarriers. The sub-carrier spacing is set to $240$ kHz to match the requirements for 6G. A single frame consists of 16 OFDM symbols per transmission interval, incorporating a Kronecker-based pilot pattern to improve channel estimation. This resource grid is transmitted by the vehicle and passes through a UMa channel as specified by 3GPP \cite{c115}. We take the AV speed range of [$60-120$]km/h for training. In the channel, the signal undergoes path loss and shadow fading. \\
At the BS, our proposed H-NR model takes the received resource grid and reconstructs the transmitted information bits. It replaces all the PHY functionalities of the BS, including resource grid demapping, channel estimation, signal equalization, demodulation, and channel decoding. Detailed parameters of the wireless communication system under consideration is listed in Table~\ref{table2}.
\begin{table}[b]
    \centering
    \caption{Wireless Communication Simulation Parameters}
    \begin{tabular}{|c|c|} \hline
        \textbf{Parameter} & \textbf{Value} \\ \hline 
        Carrier Frequency & 28GHz \\ \hline
        Physical Channel & UMa \\ \hline
        Modulation & 64 QAM \\ \hline
        Bits per symbol & 6 \\ \hline
        Code rate & 0.5 \\ \hline
        Subcarrier Spacing & 240KHz \\ \hline
        Delay Spread & 266ns \\ \hline
        Transmitter Antenna & 1 \\ \hline
        Receiver Antennas & 2 \\ \hline
        OFDM symbol & 16 \\ \hline
        Pilot Pattern & kronecker \\ \hline
        Pilot Symbol Indices & $[2, 11]$ \\ \hline
        Fast Fourier Transform Size & 129 \\ \hline
        Guard carriers & $[12, 12]$ \\ \hline
        Minimum AV Speed & 0 km/h \\ \hline
        Maximum AV Speed & 120 km/h \\ \hline
    \end{tabular}
    \label{table2}
\end{table} \\
\textbf{System Level Setup:} At the system level, we use real-world data from DeepSense 6G \cite{c16} and nuScenes \cite{c18} to test our proposed H-NR model. To cover data from all AV sensor applications, we consider image, audio, GPS, LiDAR, and radar data. Each of these modalities play a distinct and vital role in autonomous driving. For instance, image data enable object detection and traffic sign recognition, which are essential for safe navigation. LiDAR and radar provide complementary 3D spatial awareness and depth information, enhancing the vehicle’s ability to detect and track surrounding objects. This environmental awareness is crucial for real-time decision-making and path planning. Moreover, GPS data provides accurate global positioning, which is especially critical for navigating congested urban environments. Audio data, on the other hand, facilitates seamless human-vehicle interaction through voice commands and alerts, improving user experience and accessibility. \\

\section{Results and discussion} \label{s6}
In this section, we evaluate the performance of our proposed H-NR-driven end-to-end wireless communication framework against state-of-the-art methods. It is divided into two subsections. In the first subsection, we assess the performance of our proposed model by varying PHY parameters that influence the channel behavior. This helps us to analyze the behavior and adaptability of our proposed model in different scenarios. In the second subsection, we utilize multi-modal data including image, audio, GPS, radar, and LiDAR from the DeepSense 6G \cite{c16} and nuScenes \cite{c18} datasets for V2X communication. We shift the data to its bit level representation and apply PHY operations to it, including transmission, a 3GPP-defined wireless channel model, and reception, to simulate real-world communication scenarios. For comparison, we benchmark our proposed model against four state-of-the-art methods, including:

\begin{enumerate}
    \item \textbf{\textit{Perfect CSI:}} This approach assumes perfect channel knowledge and uses an ideal LMMSE equalizer to mitigate interference and noise. The equalized symbols are demodulated to compute LLRs, which are then decoded with an LDPC decoder to recover the original message. This setup represents an upper performance bound for benchmarking.

    \item \textbf{\textit{Baseline Receiver:}} It is a traditional stochastic technique used to estimate the channel response. It utilizes known pilot symbols transmitted on predefined subcarriers to infer the channel conditions. Following LS-based channel estimation, the receiver employs an LMMSE equalizer to counteract the channel effects. The equalized signals are then processed by a QAM demodulator to retrieve the LLRs. Subsequently, an LDPC channel decoder is applied to perform forward error correction and recover the original information bits. This method does not rely on prior knowledge of the channel statistics and is commonly used as a baseline in communication systems.

    \item \textbf{\textit{Neural Receiver – DeepRx:}} It is a CNN-based neural receiver, proposed in \cite{c11}. It learns to map received OFDM signal to transmitted bit sequences by replacing the stochastic channel estimation, signal equalization and QAM demodulation. After symbol prediction, a resource grid demapper extracts data symbols, which are then decoded using an LDPC decoder to recover the original information bits.

    \item \textbf{\textit{Neural Receiver – TransRx:}} It is a transformer-based neural receiver, proposed in \cite{c112}. It performs end-to-end signal detection using attention mechanism to capture long-range dependencies in OFDM signals. It maps received signals to bit-level outputs, integrating channel estimation, signal equalization and QAM demodulation within its architecture. After predicting the LLRs, a resource grid demapper extracts data symbols, and an LDPC decoder recovers the original transmitted information through error correction.
\end{enumerate}

\begin{figure}[t!]
  \centering
  \includegraphics[width=0.50\textwidth,height=0.25\textheight]{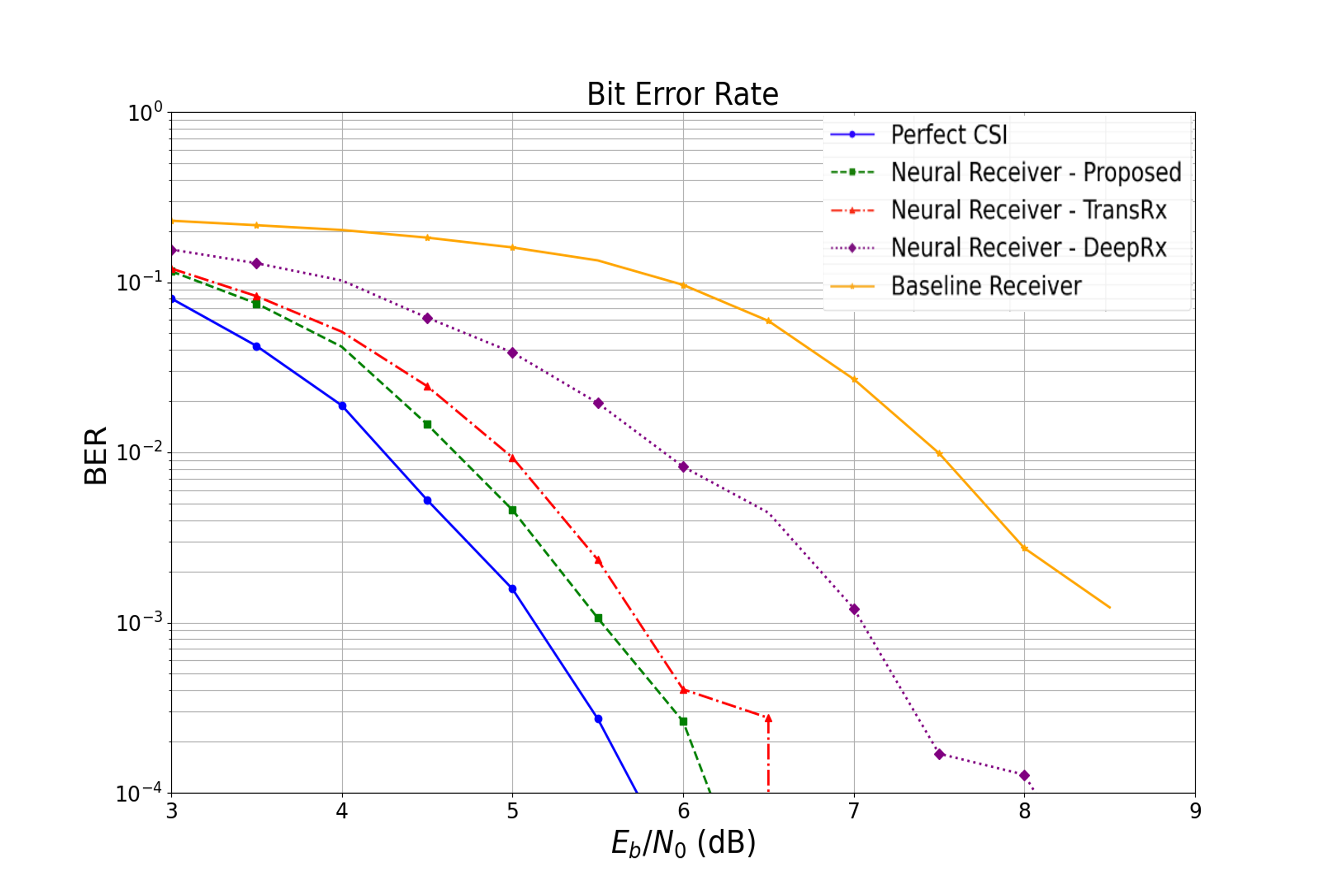}
  \caption{BER w.r.t SNR for UMa Channel Model with vehicle speed range of [$0-60$]Km/h and carrier frequency of $28$GHz}
  \label{fig7}
\end{figure}

\subsection{Channel Level Simulation: } 

We assess the performance of our proposed H-NR model based end-to-end communication framework in a low speed range of [$0-60$]km/h, with the signal transmitted over an urban channel at a frequency of $28$GHz. We compute the analysis by generating Bit Error Rate (BER) with respect to (w.r.t) Signal to Noise Ratio (SNR) graph. The results are shown in Fig.~\ref{fig7} where we observe that our proposed neural receiver model outperform state of the art methods. Specifically, our proposed neural receiver converges to minimal error rate at $6.4$dB, achieving $0.40$dB improvement compared to TransRx, which converges at $6.5$dB and a $1.9$ dB improvement compared to DeepRx, which converges at $8$dB. The baseline receiver, on the other hand, achieves similar performance at relatively higher SNR of $9.5$dB. This highlights the efficiency of the proposed H-NR receiver in achieving minimal BER at reduced SNR levels, making it highly suitable for reliable reconstruction in urban vehicular scenarios.

\begin{figure}[t!]
  \centering
  \includegraphics[width=0.50\textwidth,height=0.25\textheight]{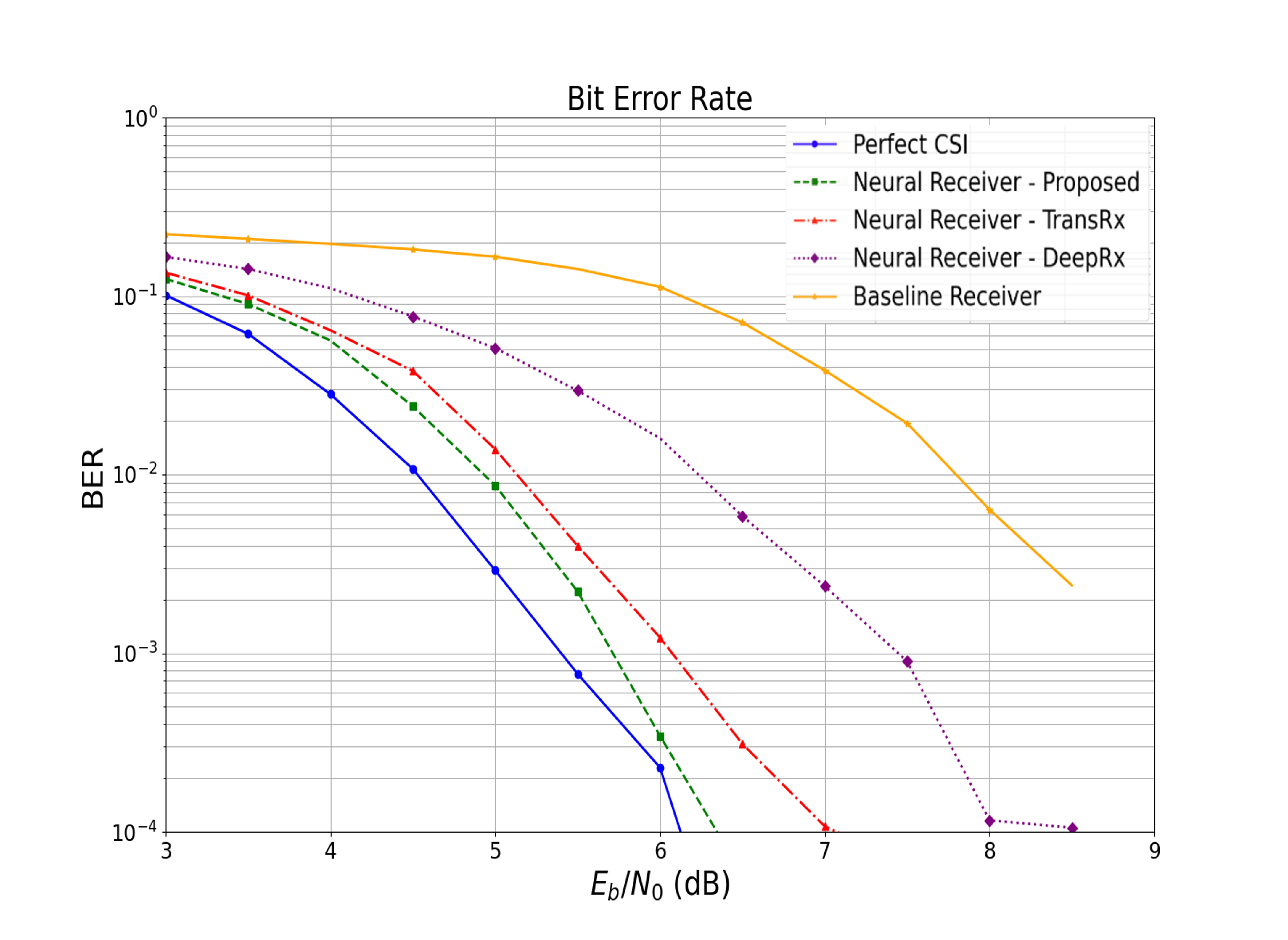}
  \caption{BER w.r.t SNR for UMa Channel Model with vehicle speed range of [$60-120$]Km/h and carrier frequency of $28$GHz}
  \label{fig5}
\end{figure}

To evaluate the performance of our proposed H-NR neural receiver in a highly mobile environment, we test the performance by increasing the AV speed ranging from [$60-120$]km/h, with signal transmitted over a UMa channel at a frequency of $28$GHz. As shown in Fig.~\ref{fig5}, our proposed neural receiver model reconstructs the transmitted signal at an SNR of approximately $6.5$dB, outperforming state-of-the-art neural receivers, TransRx and DeepRx, which require SNR of $7$ dB and $8.5$ dB, respectively. The baseline receiver, on the other hand, reaches minimal BER at an SNR of $16$dB. These results highlight the efficiency and robustness of our proposed model in high-mobility scenarios.
The significant improvement over TransRx, DeepRx, and the baseline receiver further emphasizes the model’s advantages in dynamic and challenging conditions.

\begin{figure}[t!]
  \centering
  \includegraphics[width=0.50\textwidth,height=0.25\textheight]{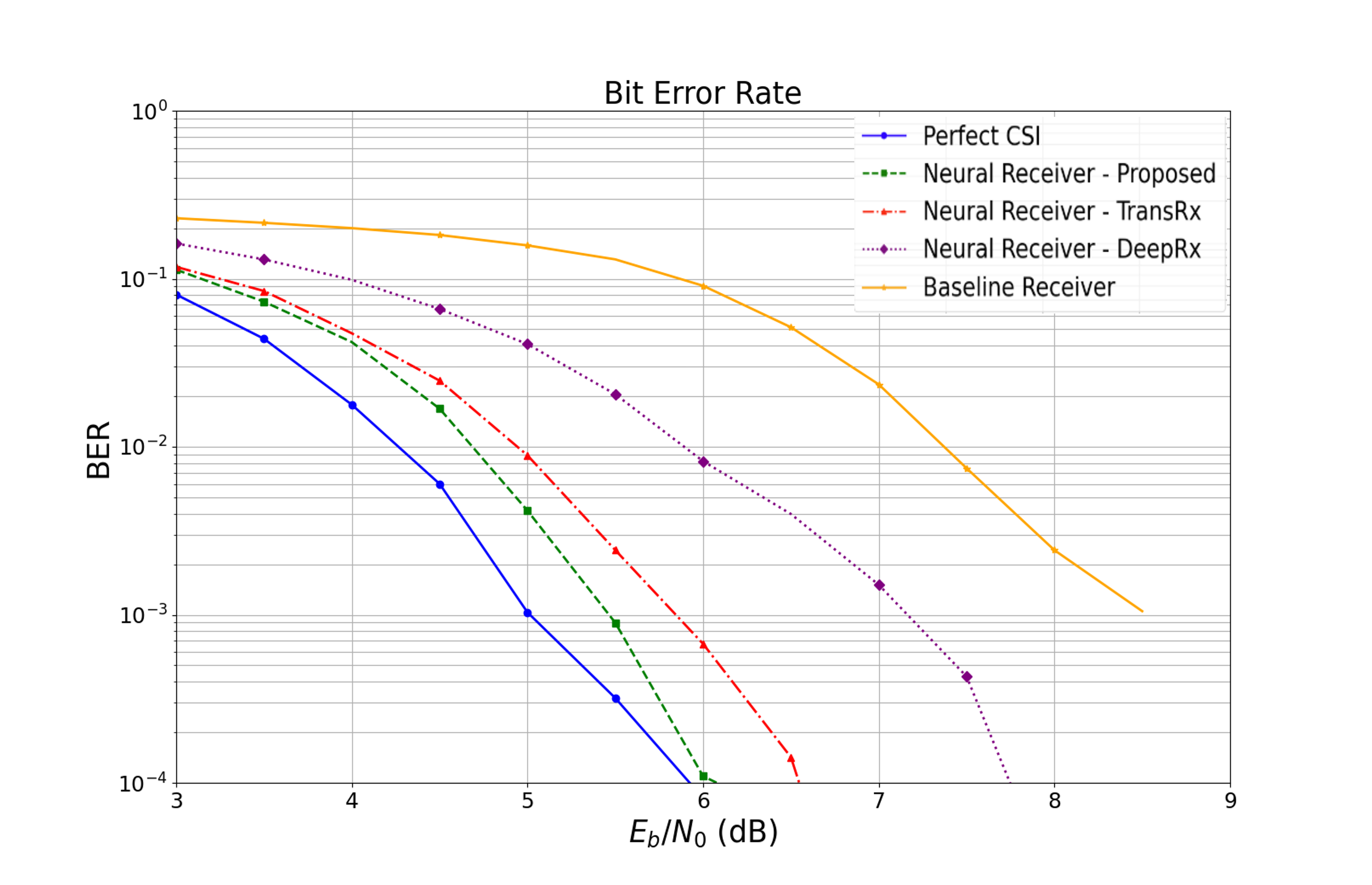}
  \caption{BER w.r.t SNR for UMa Channel Model with vehicle speed range of [$60-120$]Km/h and carrier frequency of $5.9$GHz}
  \label{fig6}
\end{figure}

To evaluate the adaptability of our proposed model for FR1 frequency band (Sub-$6$GHz) proposed by 3GPP in its release $15$ \cite{c19} for 5G-NR systems, we test it with a reduced carrier frequency of $5.9$GHz. As shown in Fig.~\ref{fig6}, our proposed H-NR model-based end-to-end framework demonstrates a notable improvement, achieving a $0.5$dB gain over TransRx, $1.75$dB improvement over DeepRx and $3.4$dB improvement over baseline receiver. These results highlight the strength and effectiveness of the proposed model, indicating its capability to ensure stable communication even in the low frequency environment.

\begin{figure}[t!]
  \centering
  \includegraphics[width=0.50\textwidth,height=0.25\textheight]{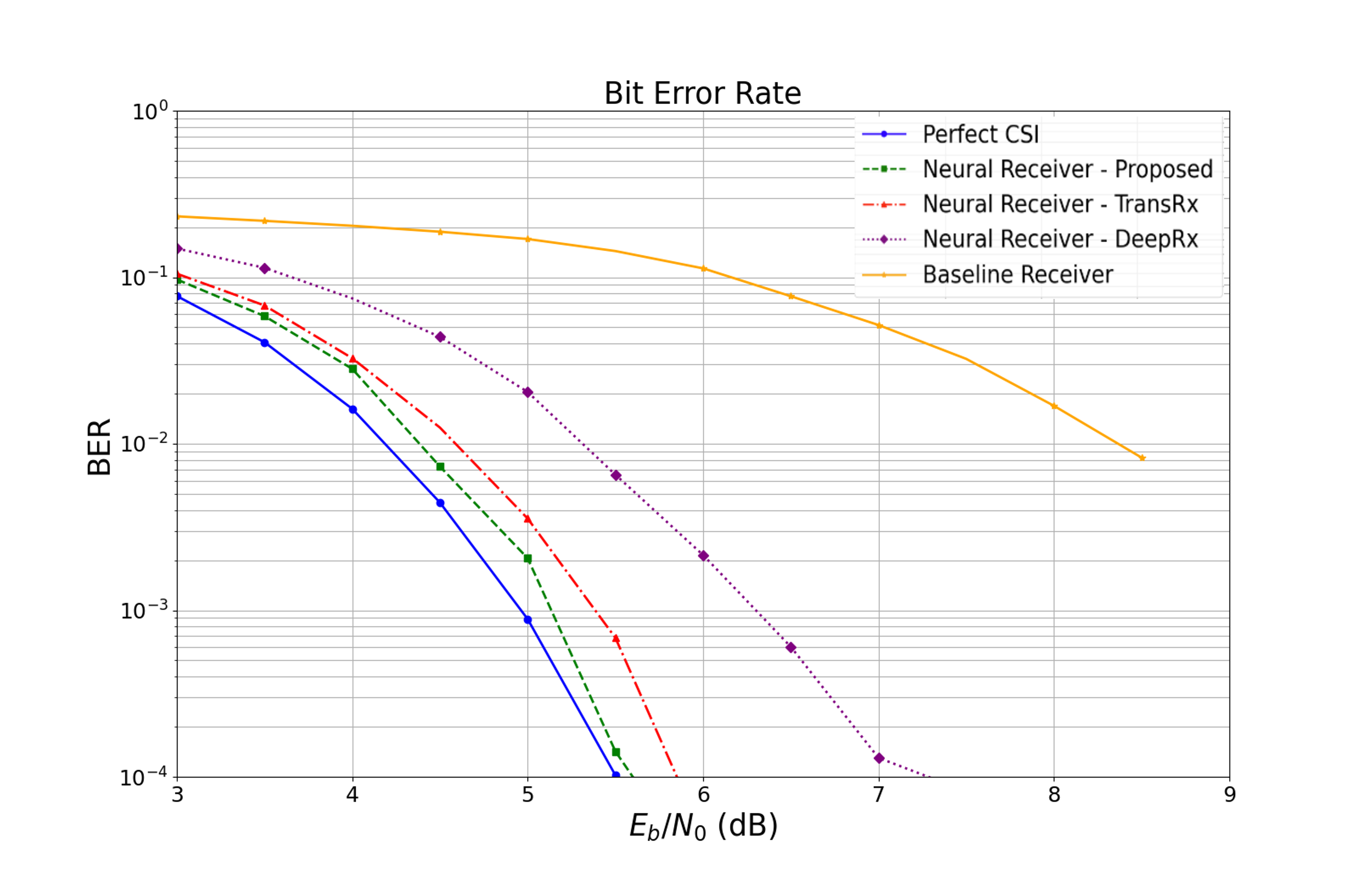}
  \caption{BER w.r.t SNR for CDL Channel Model with vehicle speed of approximately $90$Km/h and carrier frequency of $28$GHz}
  \label{fig8}
\end{figure}


To evaluate the adaptability of our proposed model for unseen channel characteristics, we test it on a different wireless channel model than the one used during training. Specifically, the signal is transmitted over a CDL channel. This shift allows us to assess how well the model adapts to unseen channel conditions. As shown in Fig.~\ref{fig8}, our proposed model outperforms state-of-the-art neural receivers, TransRx and DeepRx, as well as the baseline receiver. The proposed model achieves minimal BER by converging at an SNR of $5.60$dB. In contrast, TransRx, DeepRx and baseline receiver converges at SNRs of $5.75$dB, $7.25$dB and $12$dB respectively. These results demonstrate the strong adaptability of the proposed model, maintaining superior performance even under channel conditions unseen during training. The lower convergence SNR indicates enhanced robustness, confirming the model's potential for practical deployment in diverse wireless environments.

\begin{figure*}[t]
    \centering
    \captionsetup[subfigure]{font=footnotesize}
    \subfloat[Vehicle \\ Transmitted Image]{%
        \includegraphics[width=0.2\linewidth]{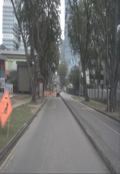}%
        \label{fig91}
    }
    \subfloat[Proposed \\ Reconstructed Image]{%
        \includegraphics[width=0.2\linewidth]{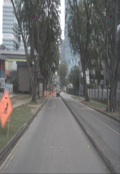}%
        \label{fig92}
    }
    \subfloat[TransRx \\ Reconstructed Image]{%
        \includegraphics[width=0.2\linewidth]{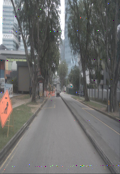}%
        \label{fig93}
    }
    \subfloat[DeepRx \\ Reconstructed Image]{%
        \includegraphics[width=0.2\linewidth]{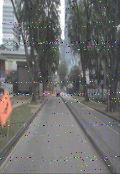}%
        \label{fig94}
    }
    \subfloat[Baseline \\ Reconstructed Image]{%
        \includegraphics[width=0.2\linewidth]{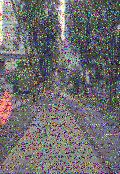}%
        \label{fig95}
    }

    \vspace{0.5em} 

    \subfloat[L1 Error Map \\ Proposed Reconstructed Image \\vs Original Transmitted Image]{%
        \includegraphics[width=0.25\linewidth]{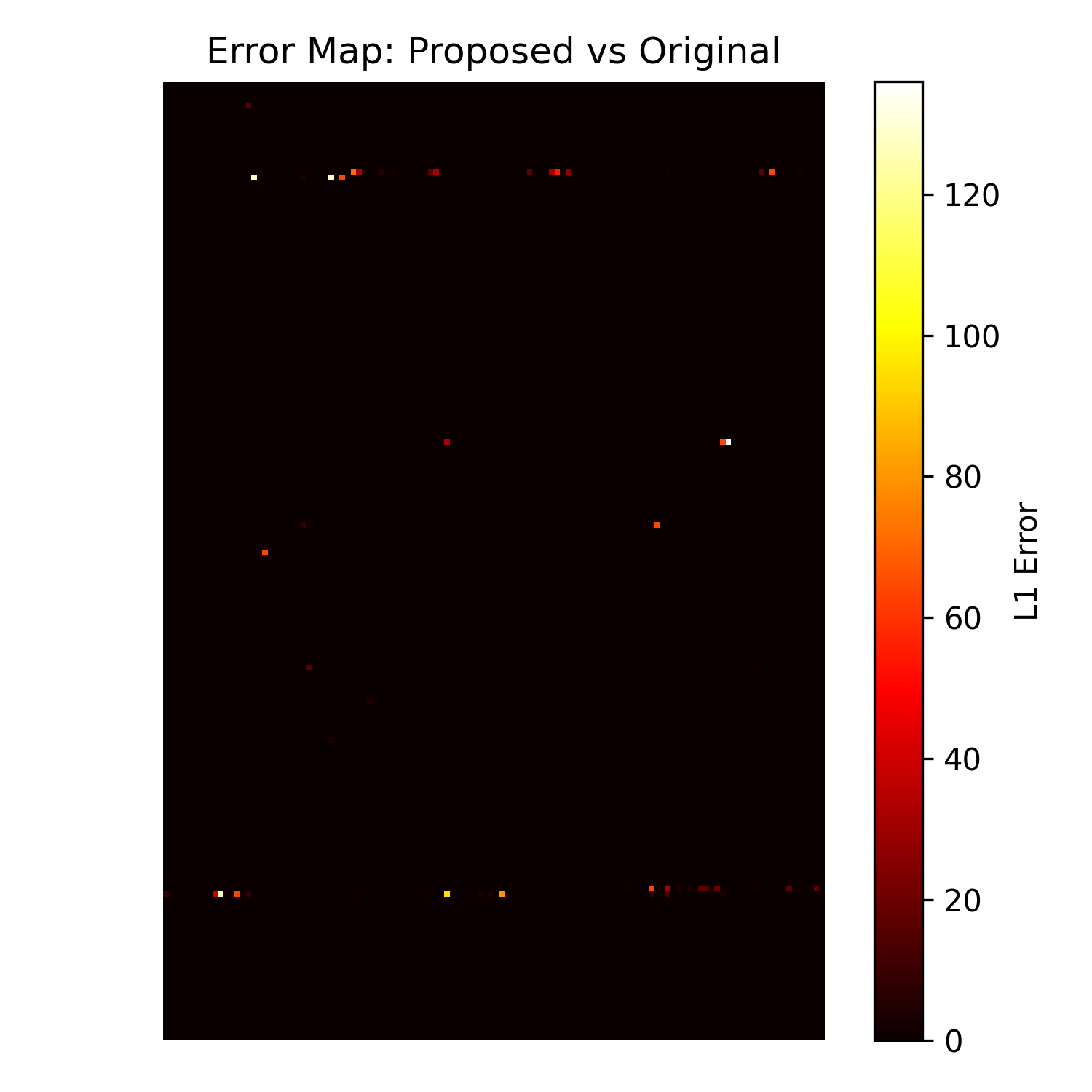}%
        \label{fig96}
    }
    \subfloat[L1 Error Map \\ TransRx Reconstructed Image \\vs Original Transmitted Image]{%
        \includegraphics[width=0.25\linewidth]{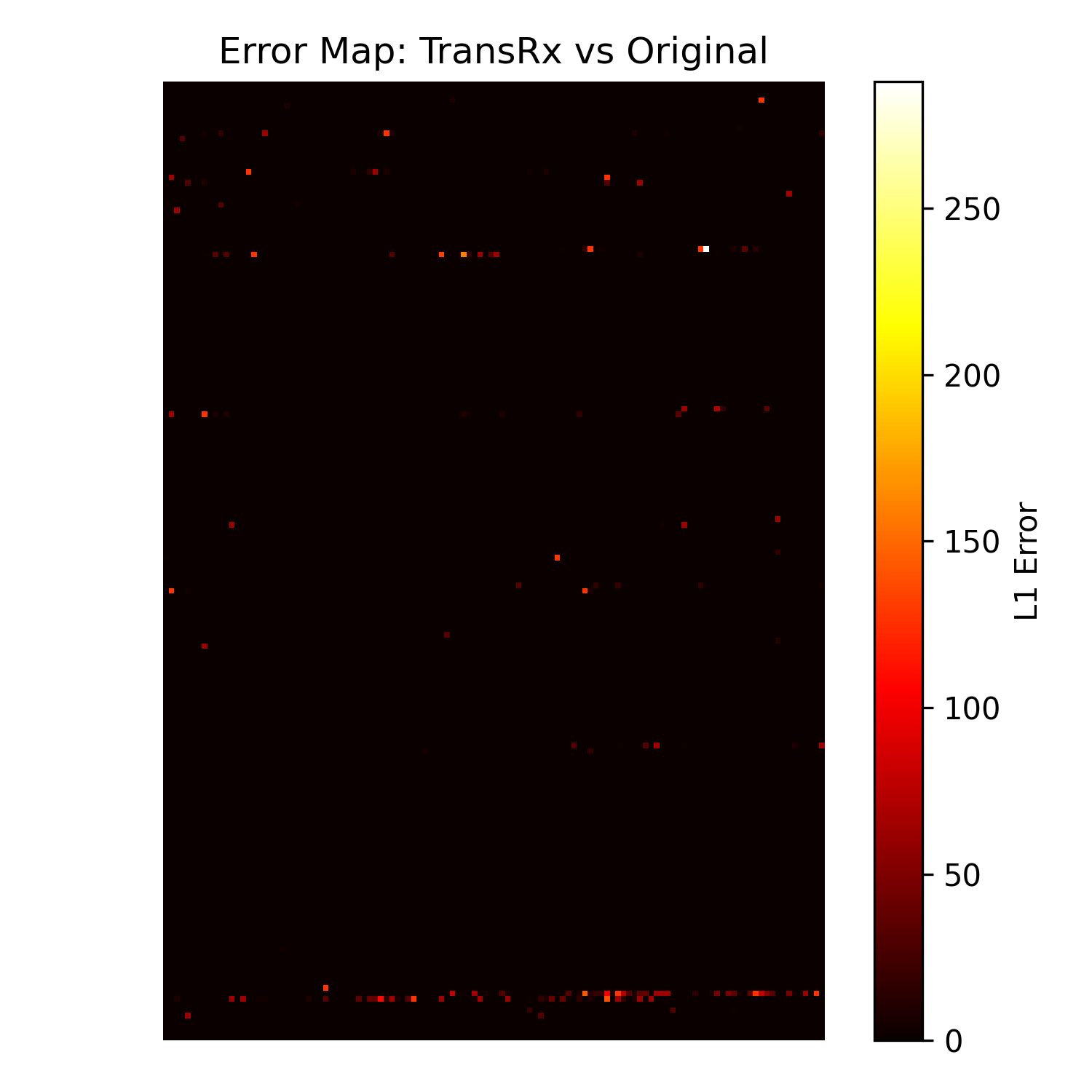}%
        \label{fig97}
    }
    \subfloat[L1 Error Map \\ DeepRx Reconstructed Image \\vs Original Transmitted Image]{%
        \includegraphics[width=0.25\linewidth]{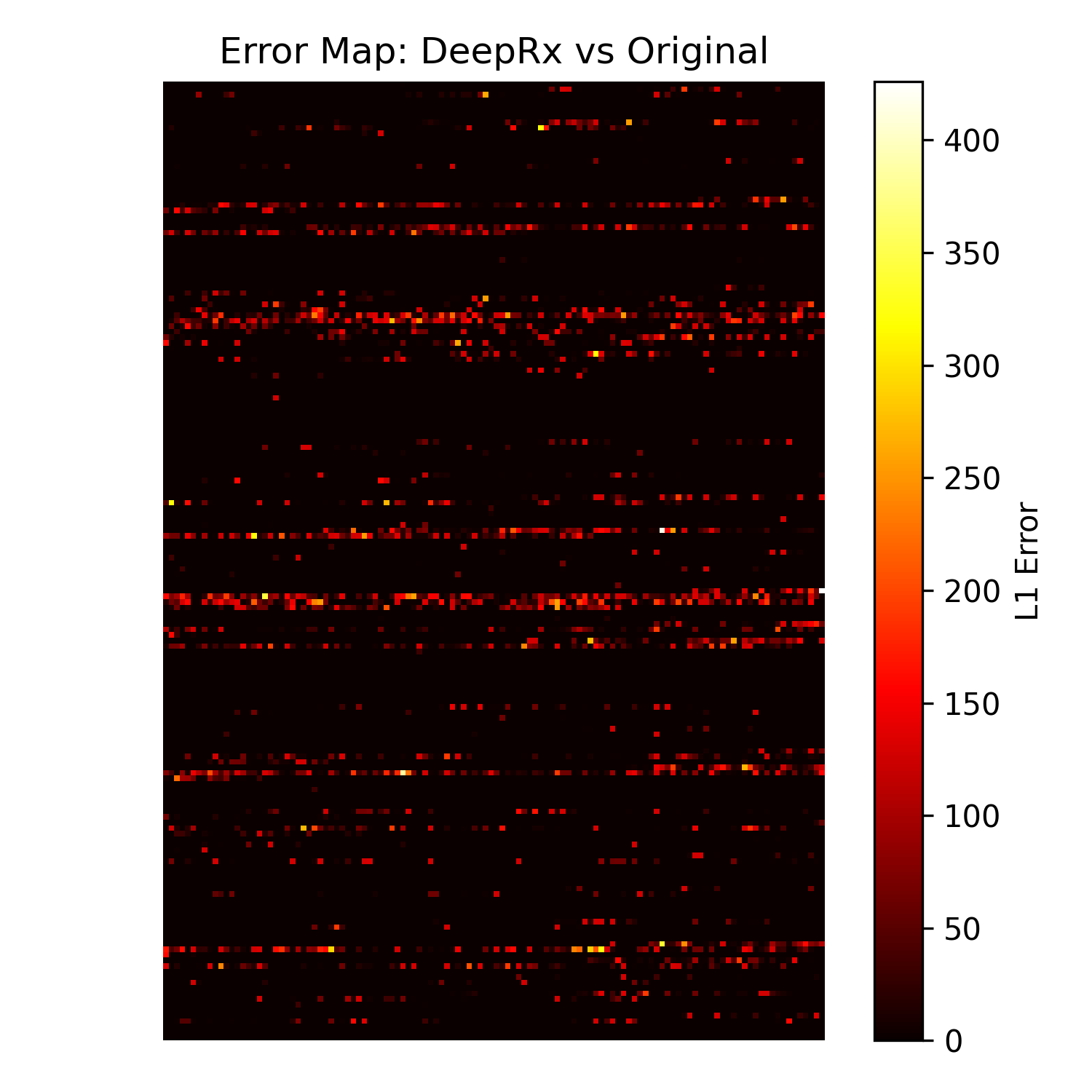}%
        \label{fig98}
    }
    \subfloat[L1 Error Map \\ Baseline Reconstructed Image \\vs Original Transmitted Image]{%
        \includegraphics[width=0.25\linewidth]{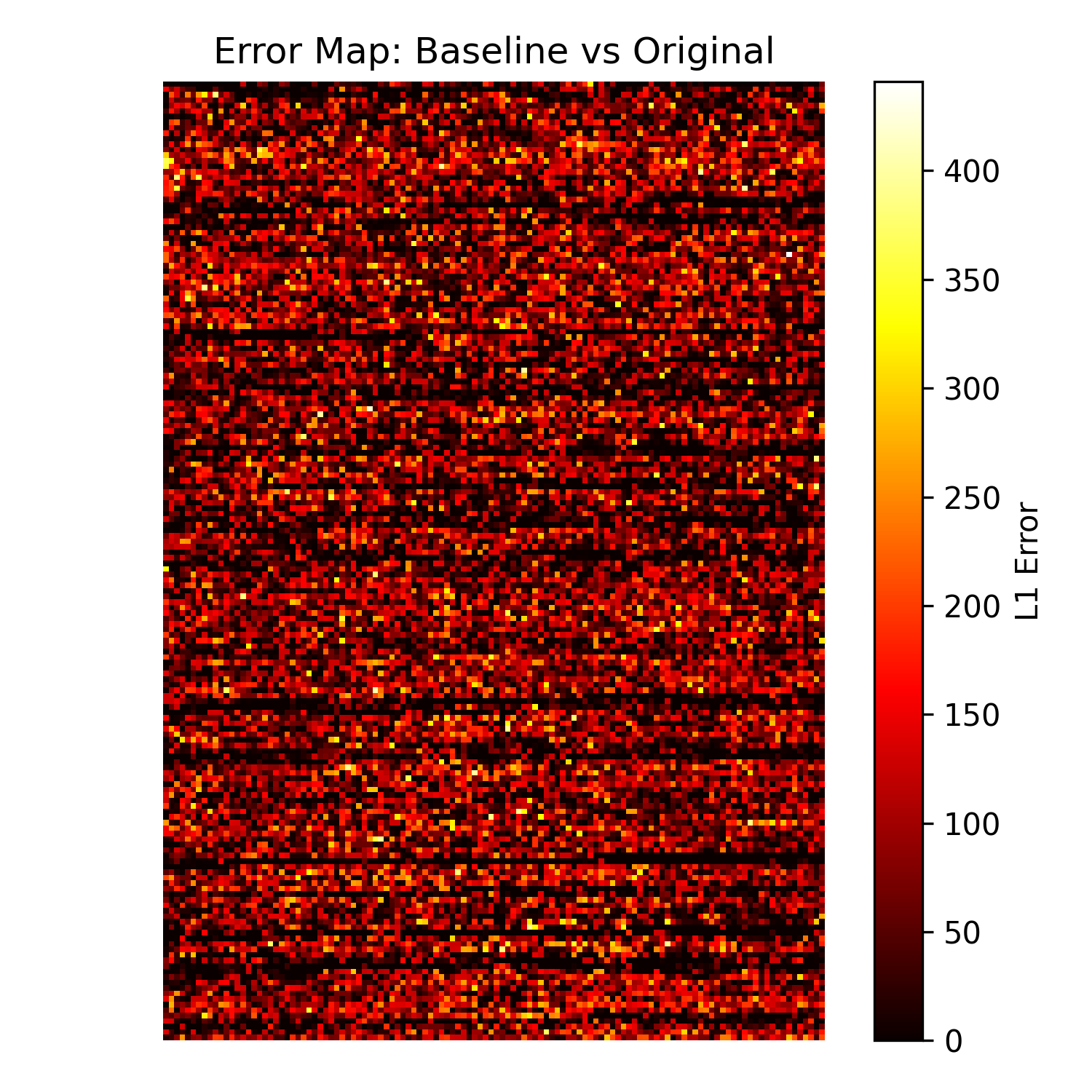}%
        \label{fig99}
    }

    \caption{Visual comparison of image reconstruction and corresponding L1 heat map analysis across different receiver architectures. The scenario is conducted over a $3$GPP UMa channel with the AV speed of $90$km/h, a carrier frequency of $28$GHz, and an SNR of $6$dB.}
    \label{fig9}
\end{figure*}

\subsection{Multi-modal Evaluation: }

We assess the performance of our proposed H-NR model-driven end-to-end wireless communication framework on multimodal data using 6G V2X communication parameters. For the testing of data transmission of all modalities, we take the vehicle speed of range [$60-120$]km/h, and the signal is transmitted over the UMa wireless channel with a carrier frequency of $28$GHz. To quantify the quality of signal reconstruction for multimodal data, we compute standard V2X parameters including:
\begin{itemize}
    \item Peak Signal to Noise Ratio (PSNR)
    \item Mean Squared Error (MSE)
    \item Root Mean Squared Error (RMSE)
\end{itemize}

\subsubsection{Image Data Transmission}
We assume that an AV captures an image and transmits it over a wireless channel with an SNR of $6$dB. At the BS, the received image is reconstructed using four different methods: our proposed H-NR neural receiver, TransRx, DeepRx, and a Baseline Receiver. As illustrated in Fig.~\ref{fig9}, our proposed neural receiver achieves the highest reconstruction quality, effectively preserving image details and significantly reducing the number of missing pixels. TransRx and DeepRx deliver moderate performance, while the baseline method performs the worst, failing to accurately recover large portions of the image. To provide a more in-depth evaluation of reconstruction quality, we employ the L1 Error Map, which visualizes the absolute pixel difference heatmaps between the reconstructed and original transmitted images, and the results are shown in Fig.~\ref{fig9}. Quantitatively, our proposed H-NR model achieves a mean L1 error of \textbf{$0.0848$}, significantly outperforming the state-of-the-art neural receivers, TransRx and DeepRx, which yield mean L1 errors of $0.3549$ and $6.8839$, respectively. In comparison, the baseline receiver records the highest mean L1 error of $78.3909$.

\begin{figure}[t!]
  \centering
  \includegraphics[width=0.50\textwidth,height=0.25\textheight]{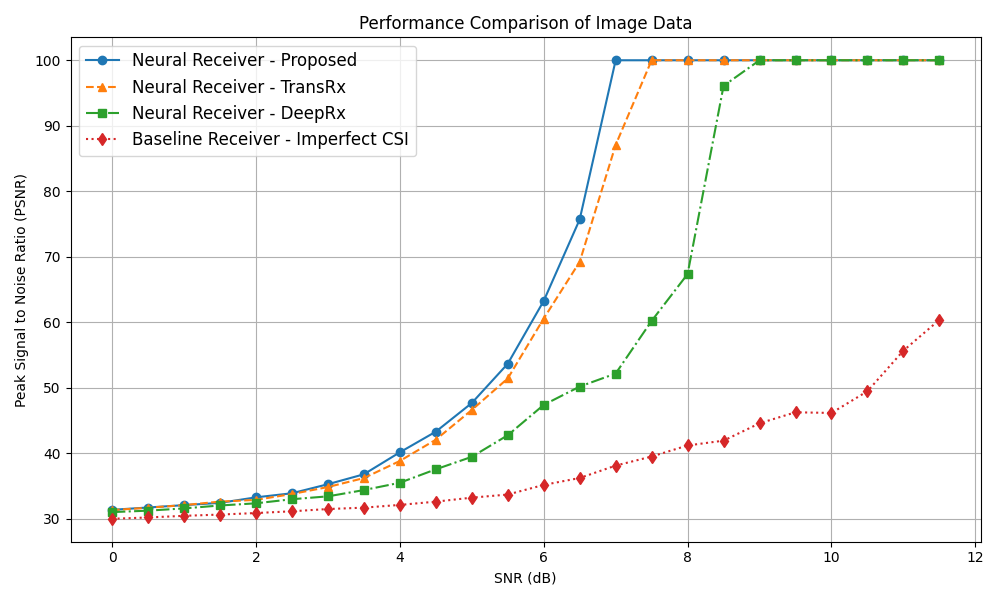}
  \caption{PSNR w.r.t SNR for image transmission over UMa Channel Model with vehicle speed range of [$60-120$]Km/h and carrier frequency of $28$GHz}
  \label{fig10}
\end{figure}

To quantitatively evaluate the reconstruction performance, we compute the PSNR as a function of the SNR. PSNR measures image quality by comparing the maximum possible signal strength to the noise introduced during the reconstruction process. As shown in Fig.~\ref{fig10}, our proposed model achieves the highest PSNR at a relatively low SNR of $7$dB. In comparison, TransRx and DeepRx achieves the similar behavior at an SNR of $7.5$dB and $9$dB. Baseline receiver, on the other hand, reach PSNR value of $56$, at a higher SNR of $11.5$dB. The highest PSNR value, such as $100$, indicates that the reconstructed image is nearly identical to the transmitted one,  indicating negligible or no distortion. 

\subsubsection{Audio Data Transmission}

\begin{figure}[t!]
  \centering
  \includegraphics[width=0.50\textwidth,height=0.25\textheight]{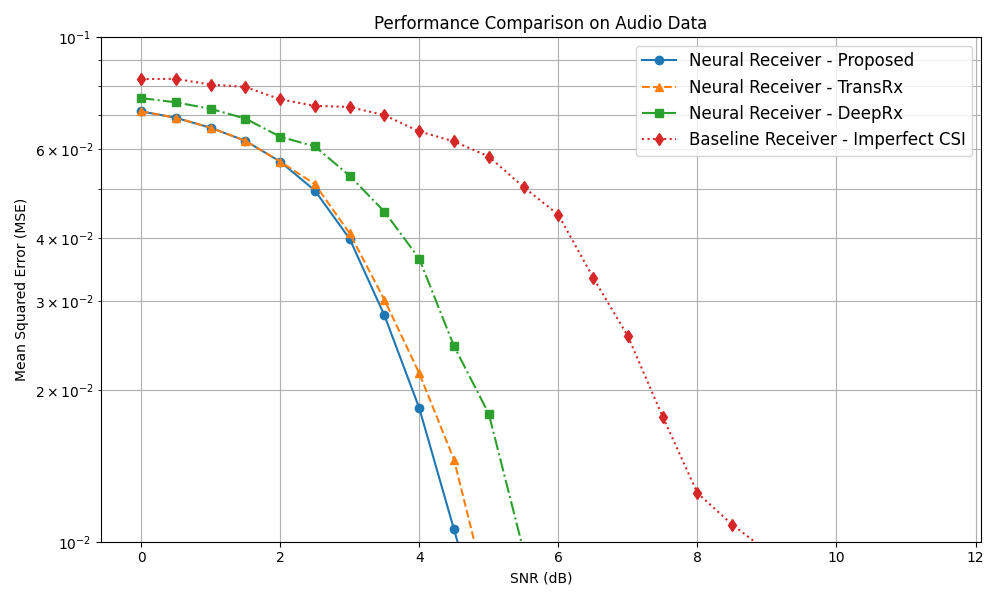}
  \caption{MSE w.r.t SNR for audio transmission over UMa Channel Model with vehicle speed range of [$60-120$]Km/h and carrier frequency of $28$GHz}
  \label{fig11}
\end{figure}

To assess the effectiveness of our proposed neural receiver in scenarios involving audio data transmission, we evaluate its reconstruction performance by computing the MSE between the original transmitted audio and the corresponding reconstructed output at the BS. We shift the Audio data into its bit level representation and append '$0$' at the end of it to match the input shape requirement of the end to end framework. Later, we perform transmitter operations on it and transmit it over a UMa channel. At the receiver, this bit level representation is reconstructed by our proposed neural receiver as well as state of the art neural receivers, TransRx and DeepRx, and baseline receiver. Once the data bits are reconstructed, we remove the padded $0$'s before converting it back to its audio representation. We conduct the analysis of all these reconstruction methods across a range of SNR values. Fig.~\ref{fig11} presents the resulting MSE values as a function of SNR. As observed, our proposed neural receiver consistently outperforms TransRx, DeepRx, and the baseline receiver architectures across all tested SNR levels. Specifically, it achieves near-zero MSE at an SNR of $4.5$dB, indicating highly accurate audio reconstruction with minimal distortion. However, TransRx, DeepRx and baseline require SNRs of approximately $5$dB, $5.5$dB and $9$dB, respectively, to reach similar reconstruction quality. 

\begin{table}[t!]
    \centering
    \caption{Minimum SNR to reach zero RMSE for GPS Reconstruction}
    \begin{tabular}{|c|c|} \hline
        \textbf{Method} & \textbf{SNR (dB)} \\ \hline
        Proposed Receiver & 6 \\ \hline
        TransRx \cite{c112} & 6.5 \\ \hline
        DeepRx \cite{c11} & 7 \\ \hline
        Baseline Receiver & 8 \\ \hline
    \end{tabular}  
    \label{table4}
\end{table}

\subsubsection{GPS Data Transmission}

We evaluate the performance of our proposed neural receiver in the context of GPS coordinate transmission. To quantify reconstruction accuracy, we calculate the RMSE between the original transmitted GPS coordinates and those reconstructed at the BS. The results are summarized in Table~\ref{table4}. As shown, our proposed neural receiver achieves the highest accuracy by successfully reconstructing the transmitted coordinates with zero RMSE at a low SNR of $6$dB. In contrast, TransRx and DeepRx achieve the same reconstruction quality at SNRs of $6.5$dB and $7$dB, respectively, while the Baseline Receiver requires an even higher SNR of $8$dB to reach zero RMSE. These results highlight the superior efficiency of our proposed model, achieving reliable coordinate recovery with an SNR gain of $0.5$ dB over TransRx, $1$ dB over DeepRx and $2$ dB over the baseline. 

\begin{figure}[t!]
  \centering
  \includegraphics[width=0.50\textwidth,height=0.25\textheight]{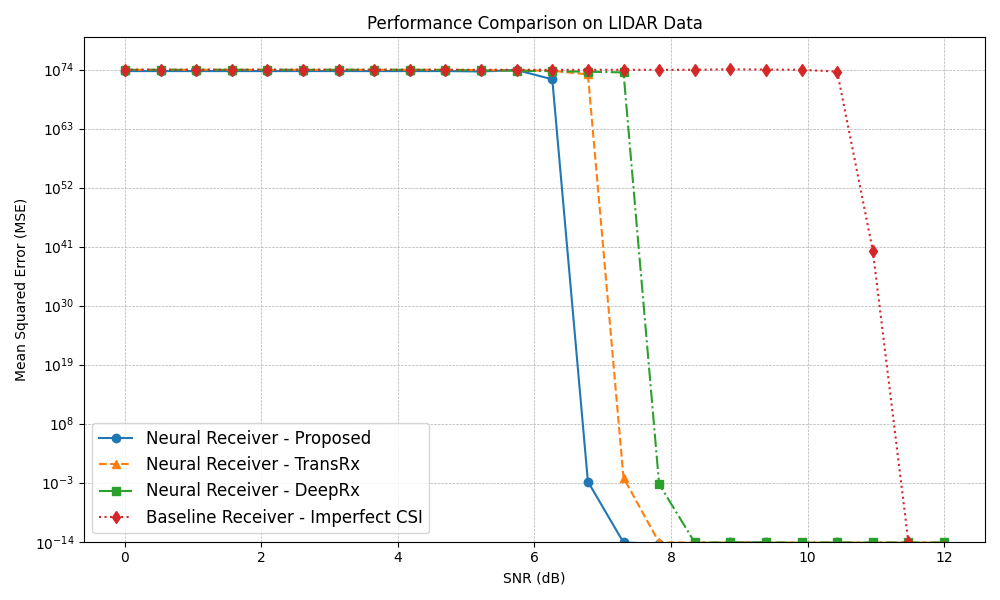}
  \caption{MSE w.r.t SNR for LiDAR transmission over UMa Channel Model with vehicle speed range of [$60-120$]Km/h and carrier frequency of $28$GHz}
  \label{fig12}
\end{figure}

\begin{figure}[t!]
  \centering
  \includegraphics[width=0.50\textwidth,height=0.25\textheight]{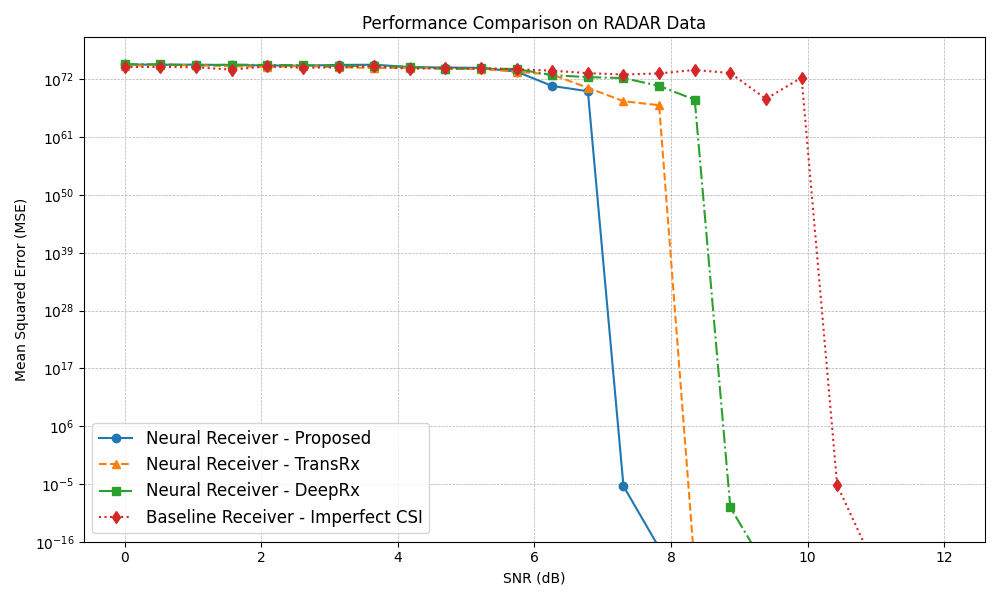}
  \caption{MSE w.r.t SNR for radar transmission over UMa Channel Model with vehicle speed range of [$60-120$]Km/h and carrier frequency of $28$GHz}
  \label{fig13}
\end{figure}

\subsubsection{LiDAR Data Transmission}

To evaluate the performance of our proposed neural receiver in the context of LiDAR point-cloud data transmission, we compute the MSE between the original transmitted point cloud and the reconstructed output at the BS across varying SNR levels. The results are illustrated in Fig.~\ref{fig12}. As shown, our proposed neural receiver significantly outperforms the state-of-the-art TransRx, DeepRx, and baseline architectures in terms of reconstruction accuracy. Our model achieves minimal MSE at an SNR of $6.75$dB. In contrast, TransRx and DeepRx require SNRs of $7.75$dB and $8.25$dB, respectively, to reach the same level of reconstruction accuracy, while the baseline lags significantly behind, achieving minimal MSE at a much higher SNR of $11.5$dB. These results clearly demonstrate the robustness and efficiency of our neural receiver, offering an SNR improvement of $1$dB over TransRx, $1.5$dB over DeepRx and $4.75$dB over the baseline.

\subsubsection{Radar Data Trasmission}

We also assess the effectiveness of our proposed neural receiver in a radar detection use case by evaluating its ability to reconstruct radar signals transmitted over a noisy wireless channel. We compute the MSE between the original transmitted radar signal and the reconstructed version across a range of SNR levels. The results are presented in Fig.~\ref{fig13}. It is observed that our proposed model achieves a near-zero MSE at an SNR of $7.75$dB. In contrast, TransRx and DeepRx achieve comparable performance at $8.25$dB and $9$dB, respectively, while the baseline method requires an SNR of $10.75$dB to achieve similar accuracy. This confirms the robustness and efficiency of our proposed neural receiver in radar signal reconstruction, which is crucial for real-time situational awareness and decision making in AV systems.

\section{conclusion and future works} \label{s7}
In this paper, we proposed a novel hybrid neural receiver (H-NR), designed to replace all PHY core functionalities of a wireless receiver, including OFDM resource grid demapping, channel estimation, signal equalization, higher order demodulation, and channel decoding. The performance of our proposed H-NR is tested across multiple scenarios, including the unseen cases during training such as a vehicle speed range of [$0-60$] km/h, a carrier frequency of $5.9$ GHz, and a CDL wireless channel. Additionally, the model is evaluated on multi-modal data, incorporating images, audio, GPS, radar, and LiDAR, to assess its adaptability to real-world applications. The results of the simulation indicate that the proposed H-NR outperforms state-of-the-art methods in terms of reconstruction and error correction quality by approximately $0.5$dB. These findings validate the effectiveness of H-NR in complex and dynamic communication settings, paving the way for intelligent, data-driven PHY design in future wireless networks for V2X communication systems.

In the future, we plan to implement a testbed for the proposed H-NR model by using USRP cards and the OpenAir Interface (OAI) platform to evaluate its performance in practical and real time communication environment.

\newpage

\vfill

\end{document}